\documentclass[aps,prl,amsmath,amssymb,twocolumn,showpacs,nofootinbib]{revtex4-1} 
\usepackage{graphicx}
\usepackage{dcolumn}
\usepackage{epstopdf} 
\usepackage{epsfig}
\usepackage{upgreek}
\usepackage{colordvi}
 \usepackage{float} 

\begin{document}

\title{Lasing on nonlinear localized waves in curved geometry}

\author{
Kou-Bin Hong$^1$, Chun-Yan Lin$^2$, Tsu-Chi Chang$^1$, Wei-Hsuan Liang$^1$,  Ying-Yu Lai$^1$,  Chien-Ming Wu$^2$, You-Lin Chuang$^{2,5}$, Tien-Chang Lu$^1$, Claudio Conti$^{3,4}$,  and Ray-Kuang Lee$^{2,5}$}

\affiliation{
$^1$Department of Photonics, National Chiao Tung University, Hsinchu 300, Taiwan \\
$^2$Institute of Photonics Technologies, National Tsing Hua University, Hsinchu 300, Taiwan\\
$^3$Institute for Complex Systems, National Research Council (ISC-CNR), Rome  00185, Italy\\
$^4$Department of Physics, University Sapienza, Rome 00185, Italy\\
$^5$Physics Division, National Center for Theoretical Sciences, Hsinchu 300, Taiwan}

\begin{abstract}
The use of geometrical constraints opens many new perspectives in photonics and in fundamental studies of nonlinear waves.
By implementing surface structures in vertical cavity surface emitting lasers as manifolds for curved space, we experimentally study the impacts of geometrical constraints  on nonlinear wave localization. We observe localized waves pinned to the maximal curvature in an elliptical-ring, and confirm the reduction in the localization length of waves by measuring  near and far field patterns, as well as the corresponding dispersion relation.
Theoretically, analyses based on a dissipative model with  a parabola curve give good agreement remarkably to experimental measurement on the transition from delocalized to localized waves.
The introduction of curved geometry allows to control and design lasing modes in the nonlinear regime.
\end{abstract}


\maketitle

Studies on the effect of geometry on wave propagation can be tracked back to Lord Rayleigh in the  early theory of sound~\cite{Rayleigh},
and extend to recent investigation as, for example, analogs of gravity in Bose-Einstein condensates~\cite{gravity-BEC}, optical event horizon in fiber solitons~\cite{optical-horizon, optical-horizon2}, Anderson localization~\cite{conti-1}, random lasing~\cite{ghof} and celestial mechanics in metamaterials with transformation optics~\cite{transformation-1, transformation-2, celestial,michinel}.
In addition to linear optics in curved space~\cite{optics-curved}, geometrical constraints affect shape preserving wave packets, including localized solitary waves~\cite{soliton-curved}, extended Airy beams~\cite{accelerating-curved}, and shock waves~\cite{wwan,conti-2}. Curvature also triggers localization by trapping the wave in extremely deformed regions~\cite{optics-curved-exp}. 

The effect of geometry on nonlinear phenomena suggests that a confining structure may alter the conditions for observing spatial or temporal solitary waves, which are due to the nonlinear compensation of linear diffraction, or dispersion effects~\cite{Yuri-book}.
In general,  self-localized nonlinear waves require an extra formation power to bifurcate from the linear modes~\cite{bifurcation}, but geometrical constraints may reduce or cancel this threshold.
In this work, we use the  vertical cavity surface emitting lasers (VCSELs) as a platform and fabricate a series of curved surfaces, including circular-ring  and elliptical-ring cavities, 
as well as a reference ``cold" device with no trapping potential. 
By introducing symmetry-breaking in geometry, i.e., passing from a circular-ring to an elliptical one,  we realize experimentally  lasing on nonlinear localized waves when the ellipticity of the ring is larger than a critical value~\cite{OL-Kuo}.
With different curved geometries, we measure the corresponding lasing characteristics both in the near and far fields,  as well as the dispersion relation, to  verify the localized modes pinned at the maximal curvature.
With state-of-the-art semiconductor technologies, microcavities in VCSELs have allowed small mode volumes and ultrahigh qualify factors, for applications from sensors, optical interconnects, printers, optical storages, and quantum chaos~\cite{VCSEL-1, VCSEL-2, nature-soliton, chaos, soliton-laser}. 
Our results clearly illustrate that with a large curvature in geometry, through an  effectively stronger confinement, one can access the  nonlinear waves easily to furnish a new generation for nonlinear-mode lasers operating at a power level comparable to linear devices.
\begin{figure}
\begin{center}
\includegraphics[width=8.4cm]{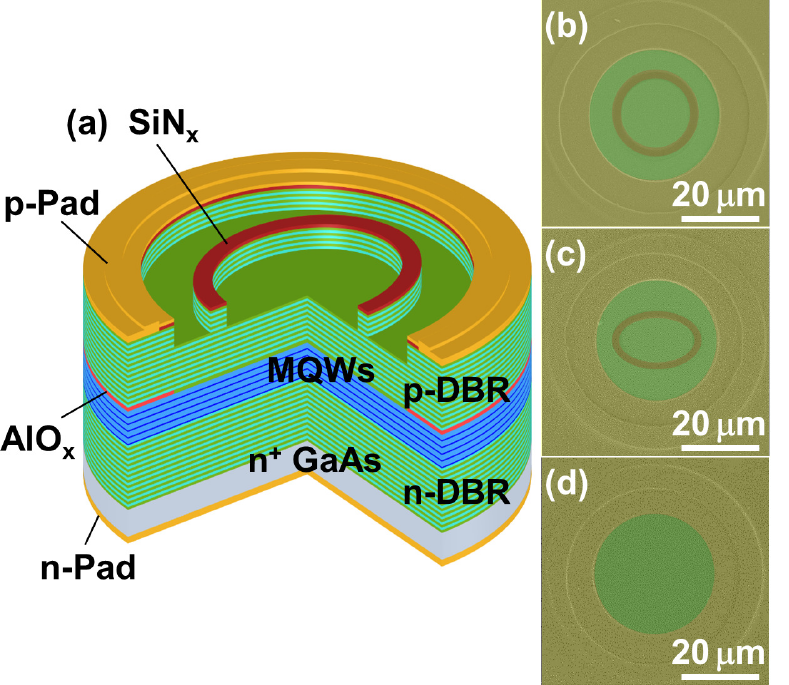}
\end{center}
\caption{(a) Schematic diagram of our electrically driven GaAs-based vertical cavity surface emitting laser (VCSEL) with a curved potential on the emission aperture defined by p-Pad metal. The emission aperture is protected by a SiN$_x$ layer with the diameter at $30$ $\mu$m. The current aperture and built-in optical confinement are defined by an AlO$_x$ layer right above the active layers. Top view of scanning electron microscope (SEM) images for (b) a circular-ring; (c) an elliptical-ring, and (d) a cold cavity used as the reference sample.}
\end{figure}

Self-focusing Kerr nonlinearity supports stationary localized waves, named  {\it crescent waves}  pinned to the boundary of a given curvature~\cite{OL-Kuo, PRL-Jisha, Kartashov2009}. 
In the case of a symmetric annular ring~\cite{PRL-Jisha}, some of us reported a random localization in the azimuthal direction, observed above a threshold formation power and bifurcating from the linear modes.  By introducing a symmetry-breaking in geometry,  as in the case of an elliptical-ring considered below, the interplay of nonlinearity and geometry modifies the localization position, localization length, and the operation current.
Here, we fabricate a series of electrically driven VCSELs with or without a predefined surface potential, as shown in Fig. 1.
The epitaxial VCSEL structure is grown on an $n^+$GaAs substrate using metal organic chemical vapor deposition systems. 
The top and bottom distributed Bragg reflectors (DBR) constructed by $17$ and $36$ pairs of Al$_{0.1}$Ga$_{0.9}$As/Al$_{0.9}$Ga$_{0.1}$As bilayers, respectively. 
The one-lambda cavity consists of $n$- and $p$-cladding layers and undoped triple multiple quantum wells (MQWs).
The thickness of quantum well was about $8$ nm and the quantum well was capped by an $8$ nm Al$_{0.3}$Ga$_{0.7}$As barrier.
Modulation doping is used in the mirror layers to reduce the differential resistance while maintaining a low free-carrier absorption loss. 
The $30$ $\mu$m (in diameter) emission window is defined by the oxide aperture (AlO$_x$); while a SiN$_x$ protection layer is deposited on $p$-DBR to  avoid the unwanted oxidation of epi-layer.
After the conventional fabrication process,  we employ the focus ion beam etching process to define the required curved structure, as shown in the Fig. 1(b) and (c).
Here, a circular-ring or  an elliptical-ring potential with a different ellipticity  is used as the curve geometry, with the etching depth at $0.6$ $\mu$m.
For the circular-ring potential, the inner and outer radius are $7$ and $8$ $\mu$m, respectively, as the top view of scanning electron microscope (SEM) image shown in Fig. 1(b).

Effectively, such a curved geometry provides a confinement along the azimuthal direction~\cite{OL-Kuo}. 
With the help of this geometrical potential, only when the ellipticty of an elliptical ring, i.e., the ratio between major and minor axes of an elliptical ring, is large enough (greater than $1.5$ in our samples), one can significantly reduce the localization length of supported solutions along the curvature, as well as its threshold power in bifurcation.
Instead, when the ellipticity is not large enough, the required nonlinearity (injection current) to have wave strongly localized is very high (not shown here). Here, To reduce the bifurcation threshold, the inner and outer radius of semi-major axis are designed as $9$ and $11$ $\mu$m, respectively; while the inner and outer radius of semi-minor axis are $5.5$ and $7.5$ $\mu$m, respectively. The corresponding ellipticity, i.e., the ratio between the major and minor axes, is $1.5$. 
The elliptical-ring provides an effective refractive index potential modulation for the localization of crescent wave. 
We also fabricate  a cold cavity, which is used as a reference sample, as shown in Fig. 1(d).

\begin{figure}
\begin{center}
\includegraphics[width=8.4cm]{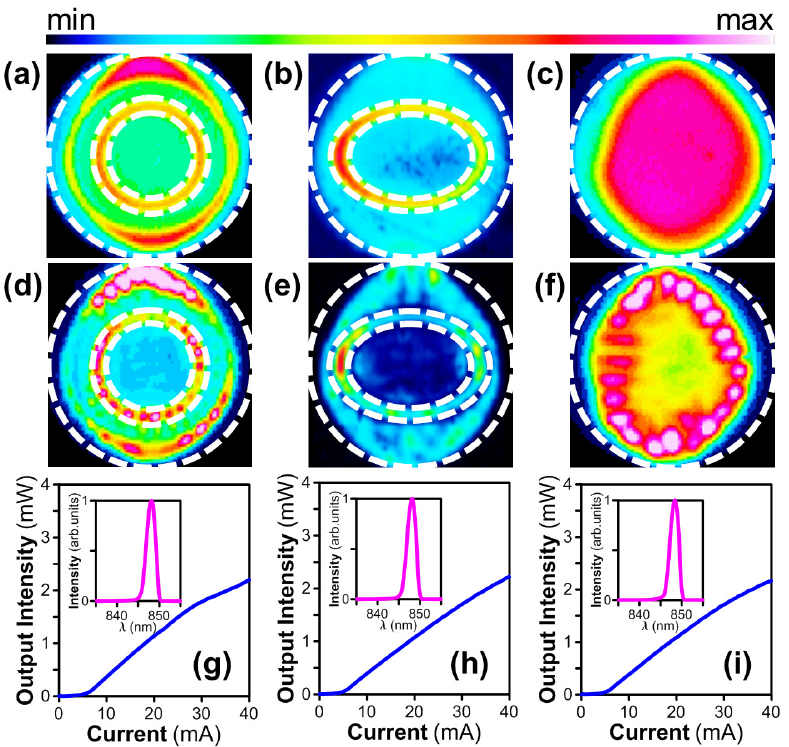}
\end{center}
\caption{Near field images of our electrically driven GaAs-based VCSELs  operated below (the first row) and above (the second row) for a circular-ring  (a, d), an elliptical-ring (b, e) potential, and the reference cold cavity (c, f), respectively.
Here, the white dashed-curves represent the edge profiles of the corresponding emission aperture defined by p-Pad metal.
Panels (g-i) show the L-I curves, output intensity versus current, for the cavities in (d-f), with lasing spectra in the insets.}
\end{figure}

At room temperature, near field images of output emission are collected into a charge-coupled device (CCD) camera using a $20X$ objective lens with a numerical aperture of $0.42$.
Figure 2(a-f) show near field  images below and above the threshold currents, i.e., the first and second rows, for the three surface potentials in Fig. 1(b-d).
Below threshold currents,  the spontaneous emission patterns follow the cavity shape, which is modulated by the surface index potential, as shown in Fig. 2(a-c), for the circular-ring, elliptical-ring, and cold cavities, respectively.
In the lasing regime, for the cold cavity, a whispering-gallery mode (WGM) is excited (Fig. 2(f)), and is usually observed in VCSEL with large aperture~\cite{VCSEL-1, VCSEL-2}.
Figure 2(d) shows the radiation pattern for the circular-ring structure composed by various spots due to the azimuthal instability~\cite{OL-YY} .

Notably, a strongly localized wave appears at the semi-major axis of elliptical-ring in Fig. 2(e), which clearly corresponds to the maximal curvature.
Here, as seen from the non-uniform radiation pattern in Fig. 2(b),  device imperfections are believed to be the reasons to cause wave localization in the left.
In addition, Fig. 2(g-i) shows the corresponding output intensity versus current,  L-I curve, revealing the absence of difference in threshold current for the elliptical-ring potential and the reference sample, i.e., about $6$ mA.
The measured lasing spectra are also shown in the insets accordingly, which reveal that all of them have the same  lasing wavelengths at $848$ nm, but with a slightly different line-widths (all smaller than $1$ nm).

\begin{figure}
\centering
\includegraphics[width=8.4cm]{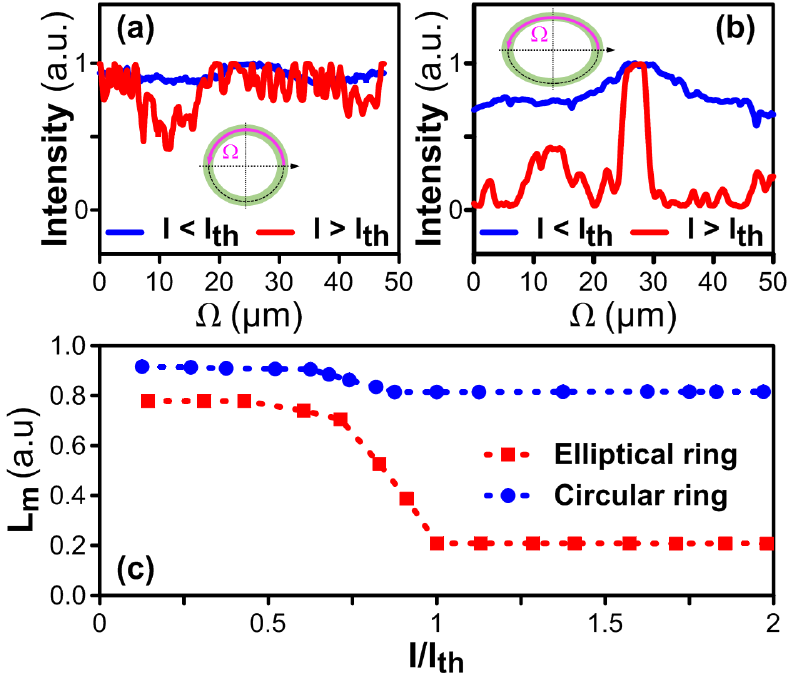}
\caption{Comparison of near field patterns for  (a)  circular-ring and (b) elliptical-ring  potentials below and above the threshold currents.
In (c), we show the localization factor $\text{L}_\text{m}$ defined in Eq. (1) as a function of the injection current $\text{I}$, normalized to the threshold current $\text{I}_\text{th}$. One clearly sees that for a circular-ring potential, the localization factor is nearly independent of $\text{I}$; on the contrary, for the elliptical-ring, the localization length changes abruptly when the injection current is larger than the threshold $\text{I}_\text{th}$.
Here, the abbreviation (a.u.) stands for arbitrary unit.}
\end{figure}

In term of the circumference, denoted as $\Omega$($\mu$m), we report the measured intensity distribution in the near field images in our electrically driven GaAs-based VCSELs in  Fig. 3(a) and (b), respectively.
For the circular-ring potential, as shown in Fig. 3(a), an almost uniform distribution can be clearly seen when operated below the threshold current $\text{I} < \text{I}_\text{th}$; while some oscillations in the intensity distribution can be found when operated above the threshold current $\text{I} > \text{I}_\text{th}$. 
As mentioned above, it is the azimuthal instability that modulates the intensity distribution along the circumference. 
However, for the elliptical-ring potential, as shown in Fig. 3(b), the intensity distribution changes from a uniform one when operated below the threshold, to a localized peak at $\Omega = 27$ $\mu$m. 
In order to give a quantitative characterization, by averaging the intensity distribution $U(\Omega)$ with the normalization to its maximum value $U_\text{max}$, we introduce a localization factor $\text{L}_\text{m}$ as the ratio to the whole circumference, i.e., 
\begin{eqnarray}
\text{L}_\text{m} \equiv \frac{1}{\text{L}_\text{C}}\oint_{\text{L}_\text{C}}  \frac{U(\Omega)}{U_\text{max}}\,d\Omega,
\end{eqnarray} 
where $\text{L}_\text{C}$ is the circumference in the curved potential.
By this definition, a  small value of localization factor means a stronger localization; while a larger localization factor, up to one,  means the wave is fully extended.
With this localization factor, $\text{L}_\text{m}$, we show in Fig. 3(c) for different normalized injection currents, $\text{I}/\text{I}_\text{th}$.
Clearly, one can see  that for a circular-ring potential, the localized factor remains almost the same, $\text{L}_\text{m} \approx 0.9$ to $0.82$, no matter we operate  at below and above the  threshold currents.
However, for the elliptical-ring, a sharp reduction in the localization factor can be found, from $\text{L}_\text{m} \approx 0.78$ ($\approx 42$ $\mu$m) to $0.21$ ($\approx 10$ $\mu$m), when the injected current is larger than the threshold condition.
This strong reduction in the localization factor demonstrates that an  elliptical-ring potential really induces the localization of lasing modes.

\begin{figure}
\begin{center}
\includegraphics[width=8.4cm]{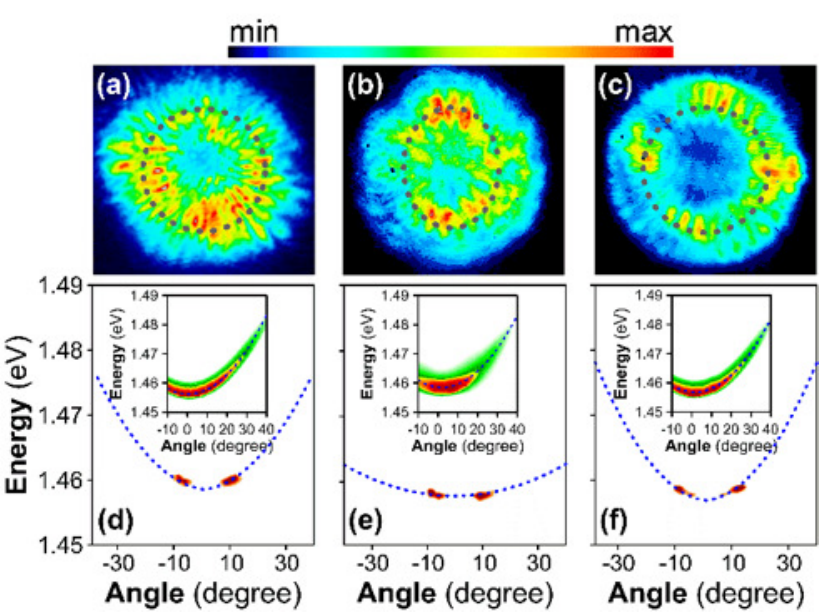}
\end{center}
\caption{Far field patterns (the first row) and  angle resolved electroluminescence (AREL) spectra measuring the energy at different angles (the second row)  for the (a) circular-ring, (b) elliptical-ring, and (c) cold cavities, operated above the threshold currents, corresponding to the near field patterns shown in Fig. 2(d-f), respectively. Insets in the second row show the corresponding AREL spectra operated below the threshold currents.}
\end{figure}

In addition to the near field intensity distribution shown in Fig. 2, we also perform the far field intensity measurements for different curved potentials.
Experiment setup for far field image measurement is similar to that of near field. 
Light radiated from the sample is focused at the Fourier plane of a $20X$ objective lens,  and an additional lens is used to project the Fourier image into the CCD. 
It is known that above the threshold currents, the corresponding far field pattern is revealed in a symmetric form both  for the circular-ring and cold cavity (WGM modes)~\cite{chaos}, as shown in Fig. 4(a) and 4(c), respectively.
On the contrary, as shown in Fig. 4(b), the far field pattern for the wave localized in the semi-major axis of an elliptical-ring becomes an elongated one,  which reflects the symmetry-breaking in the near field.

Moreover, we also perform the angle resolved electroluminescence (AREL) spectra by measuring the energy at different angles, which also gives us the corresponding dispersion relation, $E$-$k$, in the system. 
As for our AREL measurement system, it is composed by collimator, fiber (with the core radius of $600$ $\mu$m), rotary stage, and spectrometer. 
A collimator is placed $8$ cm from the sample, connected to the fiber, and rotated by using a rotary stage. 
The spectrometer has the resolution $0.7$$\textup{\AA}$, and resolution of rotation angle is $1^{\circ}$. 
The AREL spectra with different angles can be  converted into the $E-k$ curve through the expression between in-plane wavevector $k$ and rotation angle $\theta$, i.e.,  $k=(2\pi/\lambda)\sin\theta$, with the wavelength of light $\lambda$.

Below the threshold current, both the circular-ring cavity and cold cavity  have similar parabolic profiles in their $E$-$k$ curves, as shown in the insets.
Above the threshold current, as shown in Fig. 4(d) and 4(f), the output signals in the $E$-$k$ curve locate around $\pm 10^{\circ}$, but keep the parabolic profile for these two cavities.
With mode-selective mirror reflectivities, similar far field patterns located away from the center $0^{\circ}$ are also reported  in single fundamental mode AlGaAs VCSELs~\cite{mirror}.
Nevertheless, in the inset of Fig. 4(e), we reveal the $E-k$ curve for an elliptical-ring potential, which possesses a band structure with a smaller curvature at the bottom of angles within $\pm 10^{\circ}$. 
Compared to the other two cavities, a significant modification on the output dispersion relation can be seen for the elliptical-ring potential.

\begin{figure}
\begin{center}
  \includegraphics[width=8.4cm]{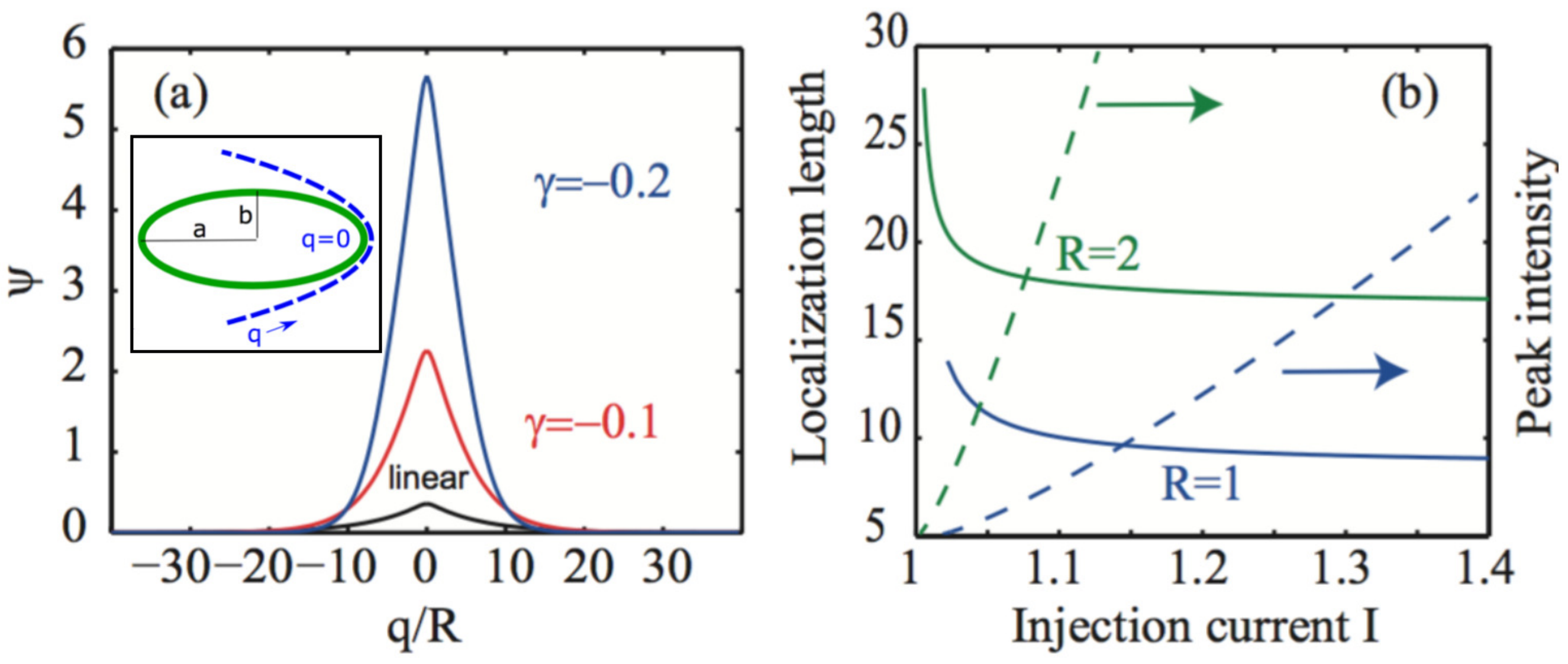}
\end{center}
\caption{\label{claudio-fig} 
(a) Profile of solutions for Eq. (4) in the linear case, and for two nonlinear cases ($\gamma = -0.1$ and $-0.2$). The inset shows the shape of an elliptical-ring (green-line) and its parabolic approximation (dashed-line) with the adopted curvilinear coordinate $q$.  
(b) Localization length (solid-line) and peak intensity (dashed-line) for the localized wave for different curvature radii versus injection current $\text{I}$ ($C=1$).
}
\end{figure}

In order to theoretically study the effect on curved manifolds on dissipartive nonlinear wave localization, we employ the curvilinear coordinate for small angles around the localization ($\Omega \cong 0$). As the dashed curve shown in the inset of Fig. 5(a), we approximate the elliptical potential shape by a parabola $k q^2$, with the curvature $k= a/2 b^2$. 
Here, the semi-major and semi-minor axes of the elliptical-ring are denoted as  $a$ and $b$, respectively.
To describe the semiconductor microcavity, we adopt the following reduced dissipative wave equation when the system reaches equilibrium~\cite{PRL-Jisha, OL-VCSEL, soliton-device}:
\begin{eqnarray}
&& \alpha \partial_t \Uppsi + \theta \Uppsi - (\alpha+i)\left[ -(1+\eta)+\frac{2C(\text{I}-1)}{1+|\Uppsi|^2} \right]\, \Uppsi \nonumber\\
&& + (\alpha-i\,{d})\,\nabla_\perp^2 \,\Uppsi  =0.
\label{eq:dnse2}
\end{eqnarray}
In Eq.~(\ref{eq:dnse2}), $\Uppsi$ gives the slowly varying complex envelope of electric field (scaled to the saturation value), $C$ is the saturable absorption coefficient scaled to the resonator transmission, ${\nabla_\perp}^2$ is the transverse Laplacian describing the diffraction in the paraxial approximation, $\eta$ is the linear absorption coefficient due to the gain material, $\theta$ is the generalized cavity detuning, and  $d$ is the diffusion constant of the carrier scaled to the diffraction coefficient. External injection current is denoted by $\text{I}$, normalized to threshold current $\text{I}_\text{th}$.
We assume  the corresponding line-width enhancement factor $\alpha$ is large enough ($\alpha \gg 1$)~\cite{a1, a2, a3}, and deal with a simplified dissipative model for the laser mode in the two transverse coordinates \cite{PRL-Jisha}
\begin{equation}
\label{eq:2dloc}
\partial_t \Uppsi =-\nabla_\perp^2 \Uppsi+ \frac{2C(\text{I}-1)}{1+|\Uppsi|^2}\,\Uppsi.
\end{equation}
We then look for a stationary solutions $\partial_t \Uppsi=0$ and transform Eq. (3) with the curvilinear coordinate $q$ along the path of curved potentials~\cite{conti-2}:
\begin{equation}
-\frac{d^2 \Uppsi}{d \eta^2}+V_G(\eta)\Uppsi= \gamma  \frac{\Uppsi}{1+|\Uppsi|^2},
\label{eq:1dlocnorm}
\end{equation}
with $\gamma=-2C(\text{I}-1)\,R^2$ serving as the``nonlinear eigenvalue'', and the radius of curvature $R$  scaled in $\eta=q/R$.
In Eq.~(\ref{eq:1dlocnorm}),  the local curvature furnishes the geometrical potential $V_G(\eta)$. Specifically, by approximating the profile of the laser by a parabola $y=k x^2$, being $k=1/2R$ and $\xi=x/R$,  we have $V_G (\eta) =-\frac{1}{4 \left[1+ \xi(\eta)^2 \right] ^3}$, along with $\eta =\frac{1}{2} \xi \sqrt{1+\xi^2}+\frac{1}{2}\sinh^{-1}\left(\xi\right)$ for $\xi(\eta)$.

The supported bound states, i.e., localized modes, can be found as the solution of  Eq. (4), which can be viewed as  a nonlinear eigenvalue equation, with the injection current $(\text{I}-1)$ embedded in $\gamma$.
With the comparison to wave packet $\psi$ in a one-dimensional potential trap, i.e., $-\psi_{xx} + V_G(x)\psi = \gamma \, \psi$,  this negative eigenvalue $\gamma = - 2C(\text{I}-1)R^2$ plays the role as the bounding energy in a trapping potential,
($ \gamma < 0$) to support a localized mode. In the linear regime,  as numerically calculated, we have the lowest eigenvalue $\gamma \cong -0.1$
with an exponentially localized ground state at $q=0$, $\psi \propto \exp(-|q| \sqrt{-\gamma})$, and localization length $L_\text{linear}\cong 4 R$~\cite{conti-2}.
Operation above the threshold current $(\text{I}-1) > 0$ is also needed to guarantee the existence of bound states, as $C > 0$.

We then consider the nonlinear regime with gain saturation, and solve Eq.~(\ref{eq:1dlocnorm}) numerically by increasing the overall energy of field.
Any bound state solution corresponds to the case: $\gamma<0$ and injection current $I=1+|\gamma|/(2C R^2)$.
One can see that the bounding energy is divided by a factor of $1+|\Uppsi|^2$, which means that the nonlinearity helps to compensate the required bond energy.
Profiles of solutions are shown in Fig. 5(a) for linear case, and for two nonlinear cases, with $\gamma = -0.1$ and $-0.2$. 
Moreover, in Fig. 5(b), we show the localization length $L_\text{loc}$ as a function of the injection current calculated by the inverse participation ratio: $L_\text{loc} (\text{I})=R\, \left(\int |\Uppsi|^2 d\eta\right)^2/\int |\Uppsi|^4 d\eta$. 
Note that the localization length reduces first and asymptotically approaches a constant value when we increase the injection current $\text{I}$ in the laser because of nonlinearity.
An increase in the injection current $I$, corresponds to a lower nonlinear eigenvalue $\gamma=-2C R^2 (I-1)$. 
One can clearly see that  the localization length is significantly reduced for a shorter radius of curvature.
By expressing quantities in real world units, we have $a=10~\mu$m and $b=6~\mu$m in our experiments.
With a comparison to the data shown in Fig. 3(c),  the localization length estimated to be reduced from   $50~\mu$m to $10~\mu$m, which is in reasonable good agreement with the experimental data (from $42~\mu$m to $10~\mu$m).

In conclusion, we have demonstrated experimentally that a proper use of geometrical constraints allows lasing on nonlinear waves at a comparable threshold current. 
Specifically, curvature and nonlinearity together pin the wave localization with a reduction in the localization length.  
The comparison of lasing modes in different curved potentials in VCSELs provides 
an extraordinary evidence of this subtle interplay between nonlinearity and geometry, and near and far field measurements support our findings. 
Our results constitute the demonstration a significant transition from delocalized to localized state by bounds and open several new ways for the excitation  of low-threshold nonlinear states and dynamics for light emission and information processing. 
Remarkable developments in this research direction not only include nonlinear optics but may involve fields as Bose-Einstein condensation and quantum optics.

\section*{Acknowledgment}
This work has been supported by Ministry of Science and Technology,  Taiwan, under Contracts NSC 102-2221-E-009-156-MY3. 
CC acknowledges support from CNR-MOST initiative and the Templeton foundation (grant number 58277). Authors acknowledge Prof. H. C. Kuo of National Chiao Tung University for measurement systems, and Prof. Y. S. Kivshar of the Australian National University for useful discussions.

\end{document}